\documentclass[aps,prx,reprint,longbibliography,nofootinbib,superscriptaddress,floatfix]{revtex4-2}

\usepackage{slashed}
\usepackage{verbatim}
\usepackage[T1]{fontenc}
\usepackage{mathbbol}
\usepackage[dvipsnames]{xcolor}
\usepackage{orcidlink}
\usepackage{ragged2e}
\usepackage[normalem]{ulem}
\usepackage[english]{babel}
\usepackage{lipsum}
\usepackage{physics}
\usepackage{dcolumn}
\usepackage{tensor}
\usepackage{comment}
\usepackage{placeins}
\usepackage{graphicx,color,overpic,mathtools}
\usepackage{amsthm,amsmath,amssymb,mathrsfs}
\usepackage{braket,bm,bbm,setspace}
\usepackage{booktabs}
\usepackage{cancel}
\usepackage{float}
\usepackage{xargs}

\usepackage{hyperref}
\hypersetup{
    colorlinks=true,
    linkcolor=RoyalBlue,
    citecolor=ForestGreen,
    urlcolor=RoyalBlue}
\definecolor{myred}{RGB}{179, 27, 27}
 
\def\be{\begin{equation}}
\def\ee{\end{equation}}
\def\bea{\begin{eqnarray}}
\def\eea{\end{eqnarray}}
\def\beq{\begin{eqnarray}}
\def\eeq{\end{eqnarray}}

\newcommand{\GSSI}{%
Gran Sasso Science Institute (GSSI), I-67100 L'Aquila, Italy}
\newcommand{\GranSasso}{%
INFN, Laboratori Nazionali del Gran Sasso, I-67100 Assergi, Italy}

\begin{document}

\title{Tidal Stripping of Matter Bound to the Secondary in Extreme Mass-Ratio Inspirals}

\author{Sreejith Nair\orcidlink{0000-0002-6428-5143}}
\email{sreejitha.nair@iucaa.in}
\affiliation{Inter University Centre for Astronomy and Astrophysics, Post Bag 4, Ganeshkhind, Pune - 411007, India}
\affiliation{Indian Institute of Technology, Gandhinagar, Gujarat - 382055, India}
\affiliation{Université Paris Cité, CNRS, Astroparticule et Cosmologie, 10 Rue Alice Domon et Léonie Duquet, F-75013 Paris, France}

\author{Sayak Datta\orcidlink{0000-0002-4774-0298}}
\email{sayak.datta@gssi.it}
\affiliation{\GSSI}
\affiliation{\GranSasso}

\begin{abstract}
Environmental studies of  extreme mass-ratio inspirals (EMRIs) have focused almost entirely on matter surrounding the primary supermassive black hole. We instead consider matter bound to the stellar-mass secondary (e.g., gas or dark matter); which can be progressively tidally stripped during the LISA-band inspiral. This changes the bound mass of the inspiraling object, modifying the gravitational-wave (GW) phase at leading order in the secondary mass. Furthermore, as the signal interpolates from an initially dressed inspiral to a nearly bare one, it can produce a characteristic inflection in the residual phase with constant mass waveform templates.
Even for an environmental mass $\sim10^{-3}\,M_{\odot}$, the cumulative dephasing relative to in band initial bound mass waveform can be larger than unity. In subsolar mass cases, the relative dephasing can reach $\mathcal{O}(10^3)\, \rm rad$.
Neglecting this effect may bias inferred EMRI parameters at the level of the fractional change in the in-band bound mass. The tidal stripping phenomena carry information about the mass and the compactness of the bound matter, enabling probes of sub-AU, planetary- to subsolar-mass environments surrounding stellar-mass black holes.
\end{abstract}

\maketitle

\noindent \textbf{\textit{Introduction.}}---The Laser Interferometer Space Antenna (LISA)~\cite{LISA:2017pwj}, and TianQin~\cite{TianQin:2015yph} will detect gravitational waves (GWs) from extreme mass-ratio inspirals (EMRIs), stellar-mass compact objects spiraling into supermassive black holes, with sufficient phase accuracy to map the spacetime of the primary and probe its astrophysical environment~\cite{ Barack:2003fp, Barausse:2014tra,LISA:2017pwj,Babak:2017tow,Gair:2017ynp,Berry:2019wgg}. This has motivated extensive investigations of environmental effects in EMRIs. Most prior analyses of such environmental effects focus on the matter bound to the primary supermassive black hole (SMBH): dark-matter spikes, gaseous accretion disks, and stellar distributions \cite{Kocsis:2011dr, Barausse:2014tra,Eda:2013gg,Eda:2014kra,Cardoso:2021wlq,Coogan:2021uqv,Cole:2022yzw,Traykova:2021dua,Vicente:2022ivh,Mitra:2025tag,Yuan:2025fde,Duque:2023seg,Brito:2023pyl,Dyson:2025dlj,Cardoso:2022whc,Cardoso:2021wlq,Fernandes:2025osu,Datta:2025ruh, Datta:2026krm}, except for Ref.~\cite{Maselli:2020zgv,Maselli:2021men,Barsanti:2022ana}, where scalar charge on the secondary was argued to have significant relevance. 

Here we introduce a fundamentally different configuration: the secondary itself carries a bound matter cloud, progressively stripped by the primary's tidal field as the binary shrinks. In contrast to all previously studied mechanisms, this modifies the \emph{bound mass} of the inspiraling secondary and therefore affects the phase evolution at leading order in secondary mass~\cite{Cutler:1994ys, Barack:2003fp}. This type of configurations remain unexplored in EMRI environmental studies, constituting a new probe of matter bound to the \emph{secondary} rather than the primary's ambient environment.
\begin{figure}[t]
\centering
\includegraphics[width=\columnwidth]{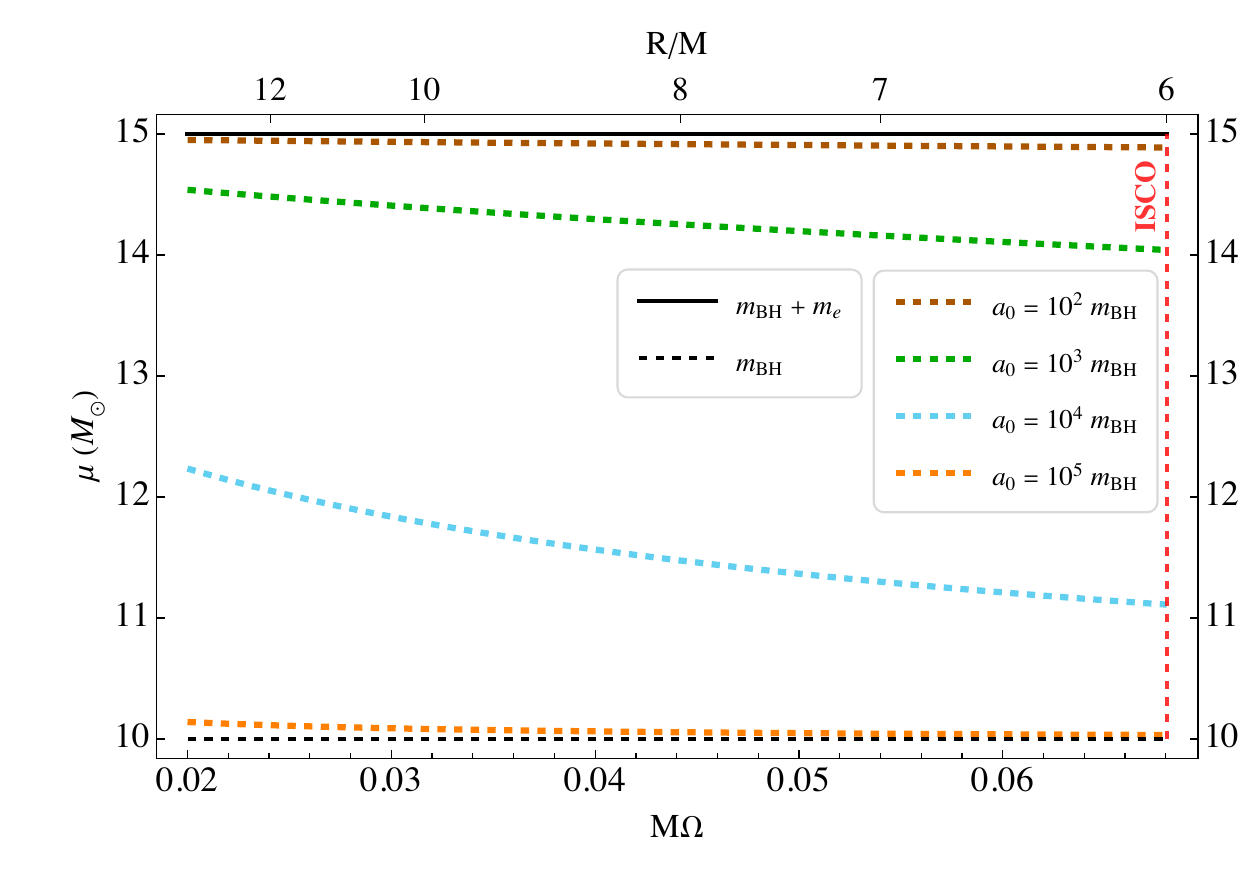}
\caption{\justifying The total secondary mass $\mu(\Omega)$, in units of $M_\odot$, as a function of the dimensionless orbital angular frequency $M\Omega$, up to ISCO. We take the primary mass $M=10^6M_\odot$, the secondary bare mass $m_{\rm BH}=10M_\odot$, and a Hernquist cloud with $m_e=5M_\odot$ and $a_0/m_{\rm BH}= 10^2,10^3,10^4,10^5$. For this primary mass, the plotted interval $0.02\leq M\Omega\leq M\Omega_{\rm ISCO}$ maps to dominant GW frequencies $f_{\rm GW}\simeq1.3$--$4.4\,{\rm mHz}$, showing that the stripping-induced variation of the bound mass is meaningful within the LISA band.}
\label{fig:mass-evolution}
\end{figure}

In this Letter, we show that this bound matter leaves a distinct imprint on the EMRI phase through tidal stripping. As the primary tidally strips material bound to secondary, the bound mass of the secondary decreases, altering the chirp rate. The signal therefore interpolates between an initially dressed inspiral and a late, nearly bare inspiral, thereby preventing any single constant-mass vacuum template from tracking the full LISA-band evolution.
This introduces a residual phase that can accumulate many radians of dephasing even for planetary- to subsolar-scale bound material. It also produces a characteristic inflection in the phase, making secondary environments both a source of EMRI parameter bias and a probe of compact-object surroundings.

\noindent \textbf{\textit{Mass model.}}---
As a diagnostic tool, we model the matter around the secondary of mass $m_{\rm BH}$ with a Hernquist profile \cite{1990ApJ...356..359H, Cardoso:2021wlq} with cloud mass $m_e$ and scale radius $a_0$, whose enclosed mass is $m_{\rm m}(r) = m_e\frac{r^2}{(r+a_0)^2}(1-\frac{2\,m_{\rm BH}}{r})^2$, where $m_{\rm BH}$ is the bare mass of the secondary, and $r$ is the separation from the center of the secondary. The Hernquist profile, is a simple analytic description of a centrally concentrated mass distribution with a finite characteristic length scale.
We do not attempt to model the detailed structure of the specific astrophysical systems. Deviations from the Hernquist profile are expected to modify the quantitative evolution of the dephasing studied below, but not the qualitative features of the stripping-induced dephasing. Three representative physical scenarios demonstrate the existence of such a configuration-- (i) a gas mini-disk within the Hill sphere of a secondary embedded in an AGN disk~\cite{Derdzinski:2018qzv,Derdzinski:2020wlw,Tagawa:2021swz, Li:2022dqh, Li:2022pnc, Chen:2023lxk, Li:2025zgo}; (ii) a dark-matter mini-spike around a primordial BH~\cite{Ricotti:2007au, Eroshenko:2016yve}; and (iii) compact bound envelope motivated by stellar collapse mass scales~\cite{1989ApJ...346..847C, Woosley:1993wj, Fryer:2011cx, Dexter:2012xk}. These scenarios are not modeled in detail here, but serve to demonstrate that the parameter space considered is plausible at the order-of-magnitude level. 

\noindent \textbf{\textit{Tidal stripping radius and bound mass.}}---Consider a primary of mass $M$ and a dressed secondary of mass $\mu$, on a circular orbit of separation $R$ around the primary, such that the enclosed mass within a radial separation $r$ from the secondary is $m_{\rm encl}(r)=m_{\rm BH}+m_{\rm m}(r)$. Material at radius $r$ from the secondary experiences the self-gravity of the enclosed mass $a_{\rm self}\sim m_{\rm encl}/r^2$ and the primary's tidal acceleration $a_{\rm tidal}\sim Mr/R^3$. Equating them gives the instantaneous Hill radius $r_H=\kappa R\left({m_{\rm encl}}/{M}\right)^{1/3},$ where $\kappa\simeq0.69$ is the Roche-limit prefactor  \cite{1983ApJ...268..368E, 2002MNRAS.329..897H}. Any material outside $r_H$ will be stripped and consequently the secondary will lose mass.

At a given orbital radius, only the material lying inside the secondary's Hill sphere stays bound to it. We therefore determine the instantaneous bound mass of the secondary as,
\begin{equation}
\label{eq:mu_omega}
\mu(\Omega)=m_{\rm encl}\!\left(r_H(\Omega)\right),\,
r_H(\Omega)=\kappa R(\Omega)\left(\frac{\mu(\Omega)}{M}\right)^{1/3},
\end{equation}
with $R(\Omega)=(M/\Omega^2)^{1/3}$. As the inspiral proceeds, $R$ and $r_H$ decrease, so material is stripped from the outside in. Thus the bound mass decreases monotonically from its initial value toward $m_{\rm BH}$, as shown in Fig.~\ref{fig:mass-evolution}, making the mass ratio $q=\mu/M$ a function of time.

\noindent \textbf{\textit{Pseudo-circular adiabatic stripping model.}}---
We model the inspiral as a sequence of instantaneous circular Schwarzschild orbits of the remnant whose bound mass $\mu(\Omega(t))$ is set by the stripping prescription in Eq.~\eqref{eq:mu_omega}.  The approximation is ``pseudo-circular'' because the orbit is treated as circular at each instant, while the mass of the secondary changes adiabatically as material is removed from its Hill sphere.  With $v\equiv (M\Omega)^{1/3}$, the specific circular-orbit binding energy is $\epsilon_{\rm orb}(\Omega)
=-1+(1-2v^2)/\sqrt{1-3v^2} <0 $.
The bound orbital energy of the dressed secondary is therefore $E_{\rm orb}=\mu(\Omega)\epsilon_{\rm orb}(\Omega)$.

The dressed secondary should be viewed as an open subsystem which is slowly losing mass as inspiral proceeds. As stripping proceeds, part of the matter previously included in $\mu(\Omega)$ leaves the Hill sphere and becomes debris.
As a first investigation, we neglect backreaction as well as model-dependent debris dynamics, including accretion, hydrodynamic torques, dynamical friction, and debris self-gravity. The surviving remnant is evolved along the instantaneous circular-orbit sequences, with the orbital energy evaluated using the instantaneous bound mass. In this pseudo-circular approximation, the waveform only gets affected through the prescribed bound mass $\mu(\Omega)$.  The inspiral is then governed by,
\begin{equation}
    \begin{aligned}
    \dot\Omega
=&
-\frac{\mathcal F_{\rm GW}}
{\mu(\Omega)\,d\epsilon_{\rm orb}/d\Omega}\,,
\\
\mathcal F_{\rm GW} =& \frac{32}{5} \left(\frac{\mu(\Omega)}{M}\right)^2 \left(M\Omega\right)^{10/3},
\label{eq:Omega_dot}
\end{aligned}
\end{equation}
where $\mathcal{F}_{\rm GW}$ is the leading quadrupole order GW flux~\cite{einstein1918gravitationswellen, 1964PhRv..136.1224P}.
The orbital phase satisfies $\dot\Phi_{\rm orb}=\Omega$, where dot represents time derivative, and we take the dominant GW phase to be $\Phi_{\rm GW}=2\Phi_{\rm orb}$ and use it to quantify the impact of evolving $\mu(\Omega)$. 

\noindent \textbf{\textit{Stripping induced dephasing}}---We numerically solve Eqs.~(\ref{eq:mu_omega}) and~(\ref{eq:Omega_dot}) for a fiducial EMRI with primary mass $M=10^6\,M_\odot$ and a dressed secondary consisting of a $m_{\rm BH}=10\,M_\odot$ black hole surrounded by a Hernquist cloud of total mass $m_e$. For each choice of $(a_0,m_e)$, we evolve the system over the final four years before ISCO. Across the representative parameter range considered here, the GW frequency at the beginning of this four-year window, $f_{\rm GW}$ is close to $1.5\,{\rm mHz}$, placing the stripping epoch well inside the LISA band~\cite{LISA:2017pwj}. 

To quantify the observable imprint of stripping, we compare the stripping signal to constant-mass reference waveforms. We denote by $\Phi_{\rm var}$ the GW phase of the stripping waveform with varying mass and by $\Phi_{\rm ref}^{(m_{\rm ref})}$ the phase of a constant-mass waveform with secondary mass $m_{\rm ref}$. The cumulative dephasing is
\begin{equation}
\delta\phi_m(t)\equiv \Phi_{\rm ref}^{(m)}(t)-\Phi_{\rm var}(t).
\end{equation}
\begin{figure}[htbp]
\centering
\includegraphics[width=\columnwidth]{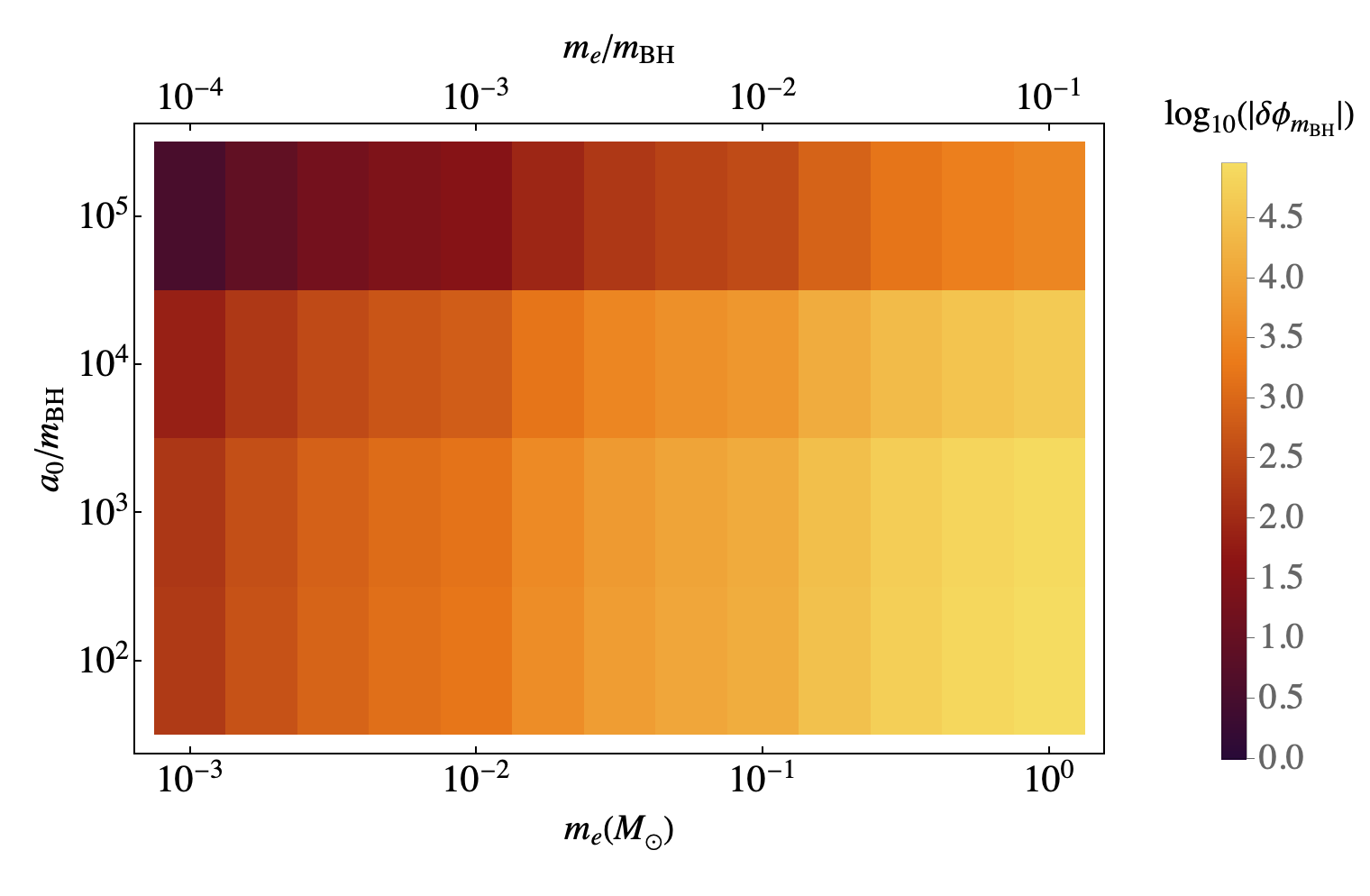}
\caption{\justifying
Four-year pre-ISCO cumulative dephasing relative to a constant bare-BH-mass template, $\left|\delta\phi_{m_{\rm BH}}\right|_{4{\rm yr}}$. The heat map shows the dependence on the total cloud mass $m_e$ and scale radius $a_0$ for $M=10^6M_\odot$ and $m_{\rm BH}=10M_\odot$.  Throughout the plotted range, $10^{-3}\,M_\odot\leq m_e\leq 1\,M_\odot$, the dephasing exceeds the nominal radian-scale threshold for waveform distinguishability.}
\label{fig:dephasing-heatmap}
\end{figure}
First, we take the reference waveform to have the bare secondary mass, $m=m_{\rm BH}$, and compute the cumulative dephasing over the final four years before ISCO across $10^{-4}m_{\rm BH}\leq m_e\leq 10^{-1}m_{\rm BH}$ and various ranges of $a_0$. The resulting values of $|\delta\phi_{m_{\rm BH}}|_{4{\rm yr}}$ are shown in Fig.~\ref{fig:dephasing-heatmap}. 
For each point in the $(m_e,a_0)$ plane, we determine the instantaneous bound mass $\mu(\Omega)$ from Eq.~(\ref{eq:mu_omega}) and integrate the corresponding stripped inspiral against the fixed-$m_{\rm BH}$ reference.  Across the region shown, the signal accumulates at least tens of radians of phase shift, exceeding $10^4\,{\rm rad}$ in the most compact and massive configurations. We find, even for the conservative environment mass $m_e=10^{-4}m_{\rm BH}$, stripping-induced dephasing is parametrically larger than the nominal radian-scale threshold for waveform distinguishability. 
This sensitivity arises because the Hill radius remains large in secondary-mass units, $r_H/m_{\rm BH}\simeq\kappa(R/M)q^{-2/3}\gg1$, 
such that the secondary retains meaningful bound mass while entering the LISA band.

For a given $m_e$ the dephasing hierarchy in Fig.~\ref{fig:dephasing-heatmap} is controlled by the ratio of the cloud scale radius $a_0$ to the Hill radius. For the fiducial four-year observation, most configurations considered here enter the band at $t_0$, four years before the observed ISCO, with dominant GW frequency $f_{\rm GW}(t_0)$ close to $1.5\,{\rm mHz}$. At this time, $r_H(t_0)\sim 10^4 m_{\rm BH}$. Clouds with $a_0> r_H(t_0)$, such as $a_0=10^5m_{\rm BH}$, are already strongly depleted through stripping before the observable inspiral begins and therefore evolve closer to the bare-$m_{\rm BH}$ template, inducing lower dephasing.
In contrast, compact clouds with $a_0\ll r_H(t_0)$, such as $a_0=10^2m_{\rm BH}$, retain most of their mass until late times. Their phase evolution therefore remains closer to a constant $\mu(t_0)$ template, while departing strongly from a bare-$m_{\rm BH}$ inspiral, despite the onset of the stripping later in the late inspiral.
For a fixed $a_0$, the dephasing increases with $m_e$. This is because the mass remaining during the in band inspiral depends directly on the total environmental mass on the secondary, when far away, which equals $m_e$. A larger $m_e$ results in a greater enclosed mass, $m_{\rm encl}$ at band entry, which is subsequently stripped during the inspiral, leading to a larger phase deviation.
In Fig.~\ref{fig:dephasing-heatmap} we find that the accumulated dephasing remains above $1\,{\rm rad}$ down to $m_e/m_{\rm BH}\sim10^{-3}$--$10^{-4}$.  For $m_{\rm BH}=10\,M_\odot$, this corresponds to $m_e\sim10^{-2}$--$10^{-3}\,M_\odot$, demonstrating sensitivity to planetary-scale bound material.  

\begin{figure}[t]
\centering
\includegraphics[width=\columnwidth]{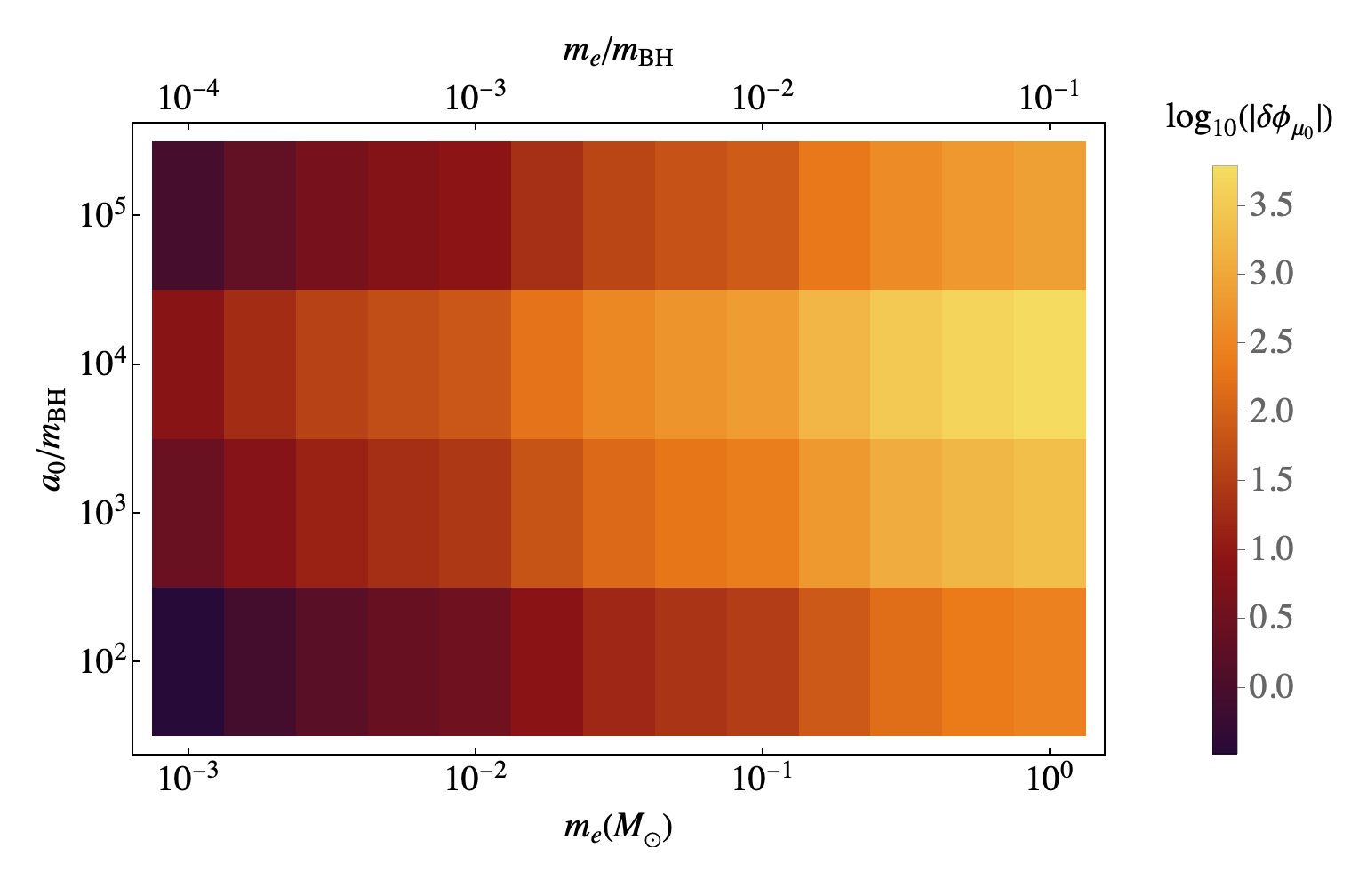}
\caption{\justifying Four-year pre-ISCO cumulative dephasing relative to a constant-$\mu_0$ (mass of the secondary four years before observed ISCO) waveform in the $(m_e,a_0)$ plane, for $M=10^6M_\odot$ and $m_{\rm BH}=10M_\odot$. Here, we kept $10^{-3}\,M_\odot\leq m_e\leq 1\,M_\odot$. The dephasing exceeds the nominal radian-scale threshold for waveform distinguishability even for planetary scale bound matter.}
\label{fig:mu0-heatmap}
\end{figure}
\noindent \textbf{\textit{Systematic bias in parameter estimation--}} A vacuum matched-filter template with a single constant secondary mass cannot reproduce the stripping waveform over the full observed band \cite{Finn:1992wt,Cutler:1994ys}. At early times, the inspiral is governed by the larger bound mass $\mu_0\equiv \mu(t_0)=m_{\rm encl}(t_0)$. At late times, after tidal stripping, the same system evolves instead as a nearly bare secondary of mass $m_{\rm BH}$, or as $m_{\rm BH}+\delta m_m$ if a residual bound envelope remains. 
The transition from the initially dressed inspiral to the stripped late-time inspiral leaves a residual mismatch that cannot be absorbed into a simple shift of the best-fit constant mass.

Fig.~\ref{fig:mu0-heatmap} illustrate the four-year cumulative pre-ISCO dephasing relative to a constant-$\mu_0$ waveform, shown in the $(m_e,a_0)$ plane for $M=10^6\,M_\odot$ and $m_{\rm BH}=10\,M_\odot$.
The total environmental mass on the secondary, when far away, $m_e$, is varied over $10^{-3}\,M_\odot \leq m_e \leq 1\,M_\odot$. The dephasing exceeds the nominal $\mathcal{O}(1)$-rad waveform-distinguishability threshold even for planetary-mass bound matter. The non-monotonic dependence on $a_0$ reflects two limiting regimes. Diffuse clouds, $a_0\gg r_H(t_0)$, are largely stripped before band entry, leaving little material to remove during the observed inspiral. Second, compact clouds with $a_0\ll r_H(t_0)$, remain mostly bound until very late times. Both extremes reduce the relative dephasing w.r.t $\mu_0$. The dephasing is therefore maximized for $a_0\sim r_H(t_0)$, where substantial bound mass survives at band entry and is efficiently stripped in band.
This could serve as a direct observational probe for the matter surrounding the secondary and if unmodeled will result in  biases in the inferred EMRI parameters.

Another consequence is a bias in the inferred time to reach ISCO.  A constant-$\mu_0$ template predicts an ISCO time $t_{\rm ISCO}^{(\mu_0)}$ that will not coincide with the observed stripping waveform, $t_{\rm ISCO}^{\rm strp}$. Stripping monotonically reduces the bound secondary mass, modifying the chirp rate. For instance, for $a_0=10^2\,m_{\rm BH}$ and $m_e=10^{-2}M_\odot$, we find an ISCO-time offset
$\Delta t_{\rm ISCO}\equiv |t_{\rm ISCO}^{(\mu_0)}-t_{\rm ISCO}^{\rm strp}|\simeq 4.6\,{\rm min}$, while for $a_0=10^3\,m_{\rm BH}$ and $m_e=10^{-3}M_\odot$ the offset remains $\simeq 3.8\,{\rm min}$. Thus, stripping affects not only the accumulated phase, but also the inferred ISCO timing.

A constant mass vacuum template would approximate the stripping signal by maximizing the LISA noise-weighted match over the observed band. Since the leading EMRI chirp rate scales linearly with the instantaneous mass of the secondary, an evolving bound mass cannot be represented by a single constant secondary mass. The natural scale of the systematic for an observation over a $\Delta t$ time window is consequently the fractional bound-mass variation in band, $\Delta\mu_{\rm obs}/\mu_0$, where $\Delta\mu_{\rm obs}\equiv\mu(t_0)-\mu(t_0+\Delta t)$. 
LISA observations of EMRIs are expected to measure the masses of the binary components with fractional accuracies as high as $10^{-5}-10^{-4}$ in favorable systems~\cite{Barack:2003fp,Babak:2017tow,Gair:2017ynp}. Thus an unmodeled stripping induced mass variation could dominate the statistical uncertainty and lead to significant parameter biases.
A quantitative analysis of the exact stripping induced systematics in EMRI parameters, along the lines of~\cite{Speri:2022upm}, requires a Fisher or Bayesian analysis, which we leave to future work.
 
\noindent\textbf{\textit{ The inflection signature and vacuum mimicry--}} The curvature of the dephasing relative to any constant mass reference template satisfies,
\begin{equation}
    \frac{d^2\delta\phi_{\rm ref}}{dt^2} \propto \dot{\Omega}_{\rm ref}-\dot{\Omega}_{\rm var}.
\end{equation}
At every $\Omega$, $\dot{\Omega}$ is proportional to the instantaneous bound mass, so the sign of $\ddot{\delta\phi}_{\rm ref}$ is controlled by $m_{\rm ref}-\mu(\Omega)$. Since stripping monotonically drives $\mu$ from $\mu_0$ toward $m_{\rm BH}$, any $m_{\rm ref}\in(m_{\rm BH},\mu_0)$ is crossed once, provided the inspiral does not reach ISCO first. Thus ${\delta\ddot\phi}_{\rm ref}$ changes sign at a unique $\Omega_*$ satisfying $\mu(\Omega_*)=m_{\rm ref}$, as long as $\Omega_*<\Omega_{\rm ISCO}$.
The appearance of an inflection depends on both the stripping history and the chosen reference template.  As shown in Supplemental Material, an analogous sign change is absent when two constant-mass vacuum templates are compared at Newtonian order. Whether this diagnostic remains unique after including the full post-Newtonian EMRI dynamics will be assessed in future works.

\noindent \textbf{\textit{Astrophysical inference from stripping--}}The stripping onset frequency $\Omega_{\rm strip}$ encodes the cloud scale radius as $a_0 \propto \Omega_{\rm strip}^{-2/3}$. Combined with the dephasing amplitude (which scales with $m_e$), this enables the possibility of joint inference of the mass and radial extent of matter bound to the secondary. 
Depending on $a_0$ and $q$, even low-mass clouds ($m_e \lesssim 10^{-3}\,M_{\odot
}$) may produce detectable dephasing, making the signal a probe for the environment surrounding the secondary.

The astrophysical inspiration comes from a secondary embedded in an AGN disk that accumulates a bound gas cloud around it~\cite{Tagawa:2021swz,Chen:2023lxk, Li:2025zgo}, remaining compact envelopes from stellar collapse as residual environment~\cite{1989ApJ...346..847C, Woosley:1993wj, Fryer:2011cx, Dexter:2012xk} and also DM mini-spikes around stellar-mass PBH secondaries ($m_e\sim 10^{-3}$--$10^{-2}\,m_{\rm BH}$, motivated by adiabatic DM accretion~\cite{Gondolo:1999ef, Bertone:2005hw, Ricotti:2007au, Eroshenko:2016yve,Kavanagh:2020cfn}). We make no detailed assumptions about these systems and treat $(m_e, a_0)$ as free parameters encoding these formation uncertainties.
A large stripping-induced phase shift signals significant bound mass; the absence can place an upper bound down to $m_e \sim 10^{-4}\,m_{\rm BH}$. LISA thus opens the GW window on sub-AU environments around stellar-mass BHs.

\noindent \textbf{\textit{Discussion--}}We identify a new class of EMRI environmental effect: tidal stripping of matter bound to the secondary, with the following central results.
First, the cumulative dephasing over a four year window exceeds the nominal $1\, \rm radian$ distinguishability threshold even for planetary scale mass bound to a $10\,M_\odot$ secondary.
Second, if stripping is left unmodeled, the evolving secondary bound mass biases inferred EMRI parameters at the level of the fractional in-band mass variation $\Delta\mu_{\rm obs}/\mu_0$, which may exceed LISA's expected measurement precision.
Third, the transition from a dressed to a nearly 
bare inspiral imprints a characteristic inflection in $\delta\ddot\phi(t)$ that cannot be reproduced by any constant-mass vacuum template at Newtonian order, providing a qualitative discriminator between stripping and vacuum evolution.
Fourth, the stripping onset frequency $\Omega_{\rm strip}$ and 
dephasing amplitude together enable joint 
inference of the mass and spatial extent of the 
bound material, opening a gravitational-wave 
window on sub-AU, planetary- to subsolar-mass 
environments around stellar-mass black holes.
Our analysis is intentionally agnostic about the 
astrophysical origin of the bound material. 
Whenever a non-negligible mass component remains 
bound to the secondary over scales comparable to 
the Hill radius, its tidal stripping during the 
inspiral generically produces observable dephasing, making this a qualitatively new environmental signature in EMRIs directly accessible to LISA.

\noindent \textbf{{Acknowledgments.}}--- We thank Biswajit Banerjee, Richard Brito, and Cristiano Ugolini for useful discussions.
S.N.\ acknowledges financial support from the ANRF-National Post Doctoral Fellowship (N-PDF) (No. PDF/2025/004999), Government of India, the Prime Minister's Research Fellowship (PMRF) (ID-1701653), Government of India, and from the Raman--Charpak Fellowship (No. IFC//4155/RCF 2024) administered by CEFIPRA. Raman--Charpak Fellowship supported his visit to APC, Paris, during which the initial parts of this project took place. S.N. thanks Luca Santoni and the members of APC for their kind hospitality during his visit. 
S.D.\ acknowledges financial support from MUR, PNRR - Missione~4 - Componente~2 -
Investimento~1.2 - finanziato dall'Unione europea - NextGenerationEU (cod.\ id.\:
SOE2024\_0000167, CUP: D13C25000660001).
The authors acknowledge CINECA for providing high-performance computing resources and support through the ISCRA initiative under project HP10CU7X29.

\bibliography{main}

\clearpage
\section*{Supplemental Material}
\begin{figure}[htbp]
\centering
\includegraphics[width=\columnwidth]{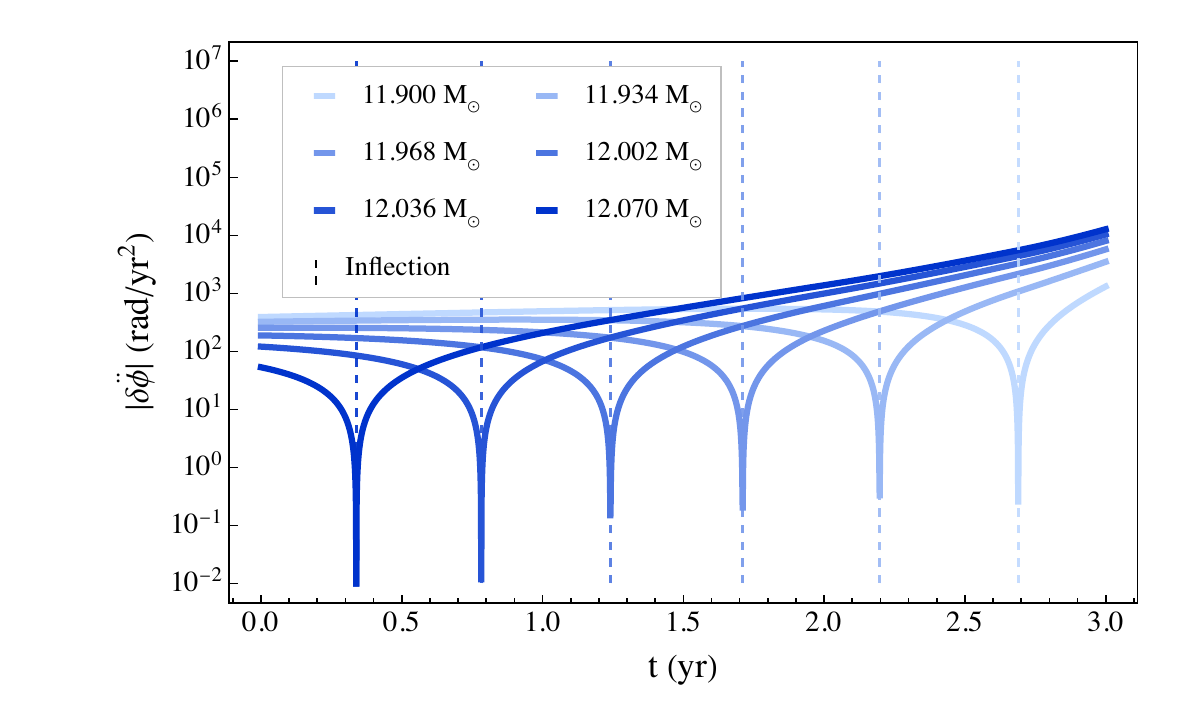}
\caption{\justifying Plot showing $|\delta\ddot\phi|$ against time in years for an inspiral starting from $1.5$ mHz, where $m_e=5~M_\odot$, $m_{BH}=10~M_{\odot}$, with $a_0=10^4 m_{BH}$, and primary mass $M=10^6~M_{\odot}$. Different curves (different shades) correspond to different constant-mass reference templates, illustrating the sign change in $\delta\ddot\phi$ that signals the inflection.}
\label{fig:inflection}
\end{figure}

\noindent \textbf{\textit{Scaling of the dephasing and inflection--}}Here we quantify the scaling of the accumulated dephasing with the
envelope mass.  For each value of $a_0$ and for each reference waveform
$m_{\rm ref}=\mu_0,m_{\rm BH}$, we fit the tabulated values in Table~\ref{tab:dephasing_extended} to a power law
\begin{equation}
    |\delta\phi| = A
    \left(\frac{m_e}{m_{\rm ref}}\right)^\beta .
\end{equation}
Equivalently, we perform a linear regression in log--log space. 
Across the eight reference curves, the best-fit power-law indices lie in the range $\beta=0.903$--$0.991$, with a coefficients of determination greater than $0.99$. Thus the accumulated dephasing is well described, over the sampled mass range, by an approximate power-law dependence on $m_e/m_{\rm ref}$.

In Fig.~\ref{fig:inflection}, we illustrate the inflection behavior of the dephasing for several reference waveforms characterized by constant template masses $m_{\rm ref}$. The quantity $|d^2\delta\phi/dt^2|$ is plotted on the vertical axis as a function of the inspiral time (in years) on the horizontal axis. The pronounced dips correspond to inflection points where $d^2\delta\phi/dt^2$ changes sign. For this example, we adopt an environmental model with $m_e=5M_\odot$ and $a_0=10^4m_{\rm BH}$.
We find that the time at which the inflection occurs decreases with increasing $m_{\rm ref}$. Physically, this is because reaching a lower reference mass requires a greater amount of stripping and therefore a longer inspiral time. The last inflection point would correspond to the lowest bound mass that can be attained through stripping and thus provides a limiting estimate of the bare compact-object mass. Consequently, by computing the inflection times for a sequence of reference masses $m_{\rm ref}$, one maybe able to infer an upper bound on the underlying bare mass $m_{\rm BH}$.
\begin{table*}
\scriptsize
\setlength{\tabcolsep}{3.0pt}
\begin{ruledtabular}
\begin{tabular}{c|cc|cc|cc|cc}
& \multicolumn{2}{c|}{$a_0=10^5\,m_\mathrm{BH}$}
& \multicolumn{2}{c|}{$a_0=10^4\,m_\mathrm{BH}$}
& \multicolumn{2}{c|}{$a_0=10^3\,m_\mathrm{BH}$}
& \multicolumn{2}{c}{$a_0=10^2\,m_\mathrm{BH}$} \\
$m_e/m_\mathrm{BH}$
& $|\delta\phi_{\mu_0}|/10^3$ & $|\delta\phi_{m_\mathrm{BH}}|/10^3$
& $|\delta\phi_{\mu_0}|/10^3$ & $|\delta\phi_{m_\mathrm{BH}}|/10^3$
& $|\delta\phi_{\mu_0}|/10^3$ & $|\delta\phi_{m_\mathrm{BH}}|/10^3$
& $|\delta\phi_{\mu_0}|/10^3$ & $|\delta\phi_{m_\mathrm{BH}}|/10^3$ \\
\hline
0.100
& 0.82 & 3.2
& 6.1 & 46
& 2.3 & 85
& 0.30 & 91 \\

0.075
& 0.62 & 2.4
& 4.7 & 36
& 1.8 & 69
& 0.23 & 75 \\

0.050
& 0.42 & 1.6
& 3.3 & 26
& 1.2 & 51
& 0.16 & 55 \\

0.025
& 0.21 & 0.84
& 1.7 & 14
& 0.64 & 29
& 0.080 & 32 \\

0.010
& 0.085 & 0.34
& 0.72 & 6.1
& 0.26 & 13
& 0.033 & 15 \\

0.0075
& 0.064 & 0.26
& 0.54 & 4.7
& 0.20 & 10
& 0.025 & 11 \\

0.005
& 0.043 & 0.17
& 0.37 & 3.2
& 0.13 & 7.1
& 0.016 & 7.9 \\

0.0025
& 0.022 & 0.087
& 0.19 & 1.6
& 0.068 & 3.7
& $8.2\times10^{-3}$ & 4.1 \\

0.001
& $8.7\times10^{-3}$ & 0.035
& 0.075 & 0.67
& 0.027 & 1.5
& $3.3\times10^{-3}$ & 1.7 \\

0.00075
& $6.5\times10^{-3}$ & 0.026
& 0.056 & 0.51
& 0.021 & 1.2
& $2.5\times10^{-3}$ & 1.3 \\

0.0005
& $4.3\times10^{-3}$ & 0.017
& 0.038 & 0.34
& 0.014 & 0.79
& $1.7\times10^{-3}$ & 0.88 \\

0.00025
& $2.2\times10^{-3}$ & $8.8\times10^{-3}$
& 0.019 & 0.17
& $6.9\times10^{-3}$ & 0.40
& $8.3\times10^{-4}$ & 0.45 \\

0.0001
& $8.7\times10^{-4}$ & $3.5\times10^{-3}$
& $7.6\times10^{-3}$ & 0.069
& $2.8\times10^{-3}$ & 0.16
& $3.3\times10^{-4}$ & 0.18 \\
\end{tabular}
\end{ruledtabular}
\caption{\justifying Cumulative dephasing $|\delta\phi_{m_{\rm ref}}|/10^3$ in units of radian for sampled $m_e/m_\mathrm{BH}<1$ values, for $a_0 = 10^2,10^3,10^4,10^5\,m_\mathrm{BH}$ and both reference waveforms $m_\mathrm{ref}=\mu_0,m_\mathrm{BH}$.}
\label{tab:dephasing_extended}
\end{table*}

\noindent \textbf{\textit{Absence of inflection for constant-mass templates--}} We adopt the standard convention where $t = 0$ denotes coalescence and the inspiral runs at $t< 0$. For two constant-mass templates with secondary masses $\mu_1$ and $\mu_2$, with $\alpha = \mu_2/\mu_1 > 1$, the leading Newtonian frequency evolution $\dot{\Omega} = \frac{96}{5}\mu M^{2/3}\Omega^{11/3}$ has the solution
\begin{equation}
    \Omega_i(t) = \Omega_0\left(\frac{-t}{\tau_i}\right)^{-3/8},
    \qquad
    \tau_i = \frac{5}{256\,\mu_i M^{2/3}\Omega_0^{8/3}},
\end{equation}
where $\tau_i > 0$ is the coalescence time from $\Omega_0$, the physical domain is $t \in (-\tau_2, 0)$, and $\tau_2 = \tau_1/\alpha < \tau_1$ since $\tau_i \propto \mu_i^{-1}$ and $\mu_2 > \mu_1$. Introducing the normalised time  $x = -t/\tau_1 \in (0, 1/\alpha)$, the frequency derivatives are
\begin{equation}
    \dot{\Omega}_i(t) \propto \mu_i\left(\frac{-t}{\tau_i}
    \right)^{-11/8},
\end{equation}
so the dephasing curvature is
\begin{equation}
    \delta\ddot{\phi} \propto 
     x^{-11/8}\,\mu_1\left(1 - \alpha^{-3/8}\right).
\end{equation}
Since $\alpha > 1$ implies $\alpha^{-3/8} < 1$, the factor $(1 - \alpha^{-3/8}) > 0$ is strictly positive and independent of $x$. Therefore $\delta\ddot{\phi}$ has constant sign and is nonzero at any physical time  $t \in (-\tau_2,0)$ for any pair of distinct constant-mass Newtonian templates  aligned at a common coalescence time. Any inflection in $\delta\ddot{\phi}_{\rm ref}$ observed in the stripped waveform is therefore a direct consequence of the time-varying secondary mass $\mu(\Omega)$ and cannot be reproduced by such constant-mass Newtonian templates. The extension to post-Newtonian corrections is left to future work.

\noindent \textbf{\textit{Backreaction of stripped material--}}
We model tidal stripping as entering the orbital dynamics solely through the evolving secondary mass $\mu(\Omega)$, neglecting any dynamical backreaction of the stripped material on the secondary's orbit. Stripping in our scenario proceeds quasi-statically: material is shed continuously from the outer edge of the Hill sphere over many orbital periods as the binary shrinks. At any instant, a stripped parcel sits at the Hill radius and is nearly co-moving with the secondary, carrying specific orbital energy and angular momentum close to that of the secondary itself. 

We therefore treat the inspiral as remaining on a quasi-circular orbit with GW-driven frequency evolution, with stripping entering only through $\mu(\Omega)$. The computed dephasing represents the contribution from mass evolution alone; any residual backreaction would introduce additional phase evolution beyond what is computed here. A full treatment of backreaction, including the angular momentum structure of the bound material and its coupling to the orbital dynamics, is deferred to future work.

\end{document}